\documentclass{aa}  
\usepackage{graphicx}

\usepackage{txfonts}

\def\zl{z$_{\rm lens}$}
\def\kms{km s$^{-1}$}

\begin{document}

   \title{COSMOGRAIL: the COSmological MOnitoring of \\
    \vspace*{1mm}  GRAvItational  Lenses III.  \thanks{Based    on
    observations made with the ESO-VLT  Unit Telescope 2 Kueyen (Cerro
    Paranal, Chile; Proposals 074.A-0563 and 075.A-0377, PI: G. Meylan).
    and on archive data taken 
    with the ESO-VLT  Unit Telescope 1 Antu (Cerro
    Paranal, Chile; Proposals 064.O-0259(A), PI: L. Wisotzki)}
    }

   \subtitle{Redshift of the lensing galaxy in 
    eight gravitationally lensed quasars}

   \titlerunning{COSMOGRAIL~III: Redshift of the lensing galaxy in 
   eight gravitationally lensed quasars}

   \author{A. Eigenbrod\inst{1} \and F. Courbin\inst{1} 
        \and G. Meylan\inst{1}
        \and C. Vuissoz\inst{1}
        \and P. Magain\inst{2}}


    \institute{
     Laboratoire d'Astrophysique, Ecole Polytechnique F\'ed\'erale
     de Lausanne (EPFL), Observatoire, 1290 Sauverny, Switzerland
     \and
     Institut   d'Astrophysique et  de  G\'eophysique, Universit\'e de
     Li\`ege, All\'ee du 6 ao\^ut 17, Sart-Tilman, Bat. B5C, B-4000 Li\`ege,
     Belgium}

   \date{Received ... ; accepted ...}

 
  \abstract
   {}
   {We measure the redshift of  the lensing galaxy in eight gravitationally lensed
    quasars in view of determining the Hubble parameter H$_0$ from the time delay method. }
   {Deep VLT/FORS1 spectra of lensed quasars are spatially deconvolved in order to
    separate the spectrum of the lensing galaxies from the glare of the much brighter
    quasar images. A new observing strategy is devised. It involves observations
    in Multi-Object-Spectroscopy (MOS) which allows the simultaneous observation of the 
    target and of several PSF and flux calibration stars. The advantage of this method 
    over traditional long-slit observations is a much more reliable extraction and 
    flux calibration of the spectra.}
   {For the first time we measure  the redshift of the lensing galaxy in three 
    multiply-imaged quasars:  SDSS~J1138$+$0314 (\zl$=0.445$),
    SDSS~J1226$-$0006 (\zl$=0.517$), SDSS~J1335$+$0118 (\zl$=0.440$), and we give a
    tentative estimate of the redshift of the lensing galaxy in 
    Q~1355$-$2257 (\zl$=0.701$). We confirm four previously measured 
    redshifts: HE~0047$-$1756 (\zl$=0.407$), HE~0230$-$2130 (\zl$=0.523$), 
    HE~0435$-$1223 (\zl$=0.454$) and WFI~J2033$-$4723 (\zl$=0.661$). In
    addition, we determine the redshift of the second lensing galaxy in 
       HE~0230$-$2130 (\zl$=0.526$). The spectra of all lens galaxies are 
    typical for early-type galaxies, except for the second lensing galaxy in
    HE~0230$-$2130 which displays prominent [OII] emission.} 
   {}


   \maketitle

\section{Introduction}

Gravitationally lensed quasars have become a truly efficient source of
new astrophysical applications, since  the discovery of the first case
by Walsh et al. (\cite{walsh79}).  These objects are potentially useful 
for measuring
the Hubble parameter H$_0$ from  the time  delay between their  lensed
images   (Refsdal \cite{Refsdal64}),  assuming  a  model  for the mass
distribution in the lensing galaxy. Conversely, for an assumed or
measured value of  H$_0$,  the mass  distribution  in the lens  can be
reconstructed  from the time delay  measurement.  The  study of lensed
quasars  is  therefore a  ``no-lose'' game,   either because  of  its
cosmological implications (H$_0$) or for the study of galaxy formation
and  evolution through the  determination  of  detailed mass maps  for
lensing galaxies, especially their dark matter content.

The  second  most important  application   of quasar lensing  involves
microlensing   by stars in   the  lensing galaxy.  Microlenses produce
chromatic magnification events in the light curve of the images of the
source, as they cross their line of sight. The amplitude, duration and
frequency of the events depend on the transverse  velocity of stars in
the  lensing galaxy, on  their  surface density,  and on the  relative
sizes of the microlensing caustics with respect to  the regions of the
source  affected by microlensing.    Photometric monitoring  of lensed
quasars in several bands or, better, spectrophotometric monitoring can
therefore yield constraints on the  energy profile of quasar accretion
disks and on the  size of the  various emission line regions (e.g.
Agol \& Krolik~\cite{agol}, Mineshige \& Yonehara \cite{mineshige}, 
Abajas et al.~\cite{abajas}).
  
All the above applications of  quasar lensing require the knowledge of
the redshift of the lensing  galaxy, which is frequently hidden  in
the glare of the  quasar images.   The present
paper is  part of a  larger effort to  carry out long term photometric
monitoring  of lensed quasars   in  the  context of  COSMOGRAIL  (e.g.
Eigenbrod  et al.~\cite{eigenbrod}). In  this paper  we focus    on the
determination of the  redshifts  of  the lensing galaxies  in  several
gravitationally lensed quasars, using deep spectra obtained with the ESO
Very Large Telescope (VLT). 
The targets were simply selected in function of
their visibility during the period of observation.

\begin{table}[t!]
\caption[]{Journal of the observations of HE~0047$-$1756. Grism and filter: G300V$+$GG435.
HR collimator: 0.1\arcsec\ per pixel. Slitlets width: $1.0\arcsec$ (R = 210 at 5900 \AA).}
\label{refer_HE0047}
\begin{flushleft}
\begin{tabular}{cccccc}
\hline 
\hline 
ID & Date & Seeing $[\arcsec]$ & Airmass & Weather \\
\hline 
1 & 18/07/2005 & 0.49 & 1.281 & Photometric\\
2 & 18/07/2005 & 0.53 & 1.191 & Photometric\\
\hline 
\end{tabular}
\end{flushleft}
\end{table}

\begin{table}[t!]
\caption[]{Journal of the observations of HE~0230$-$2130. Grism and filter: G600R$+$GG435.
SR collimator: 0.2\arcsec\ per pixel. Slitlets width: $0.5\arcsec$ (R = 1910 at 6200 \AA).}
\label{refer_HE0230}
\begin{flushleft}
\begin{tabular}{cccccc}
\hline 
\hline 
ID & Date & Seeing $[\arcsec]$ & Airmass & Weather \\
\hline 
1 & 15/12/2004 & 0.60 & 1.166 & Light clouds\\
2 & 15/12/2004 & 0.58 & 1.248 & Light clouds\\
3 & 01/03/2005 & 0.78 & 1.800 & Photometric \\
\hline 
\end{tabular}
\end{flushleft}
\end{table}
  
\begin{table}[t!]
\caption[]{Journal of the observations for HE~0435$-$1223. Grism and filter: G300V$+$GG435.
HR collimator: 0.1\arcsec\ per pixel. Slitlets width: $1.0\arcsec$ (R = 210 at 5900 \AA).}
\label{refer_HE0435}
\begin{flushleft}
\begin{tabular}{cccccc}
\hline 
\hline 
ID & Date & Seeing $[\arcsec]$ & Airmass & Weather \\
\hline 
1 & 11/10/2004 & 0.47 & 1.024 & Photometric\\
2 & 11/10/2004 & 0.45 & 1.028 & Photometric\\
3 & 12/10/2004 & 0.46 & 1.024 & Photometric\\
4 & 12/10/2004 & 0.53 & 1.028 & Photometric\\
5 & 11/11/2004 & 0.57 & 1.093 & Photometric\\
6 & 11/11/2004 & 0.56 & 1.145 & Photometric\\
\hline 
\end{tabular}
\end{flushleft}
\end{table}

\begin{table}[t!]
\caption[]{Journal of the observations for SDSS~J1138$+$0314. Grism and filter: G300V$+$GG435.
HR collimator: 0.1\arcsec\ per pixel. Slitlets width: $1.0\arcsec$ (R = 210 at 5900 \AA).}
\label{refer_J1138}
\begin{flushleft}
\begin{tabular}{cccccc}
\hline 
\hline 
ID & Date & Seeing $[\arcsec]$ & Airmass & Weather \\
\hline 
1 & 10/05/2005 & 0.82 & 1.191 & Photometric\\
2 & 11/05/2005 & 0.70 & 1.155 & Photometric\\
3 & 11/05/2005 & 0.67 & 1.133 & Photometric\\
4 & 11/05/2005 & 0.66 & 1.132 & Photometric\\
5 & 11/05/2005 & 0.64 & 1.148 & Photometric\\
\hline 
\end{tabular}
\end{flushleft}
\end{table}

\begin{table}[t!]
\caption[]{Journal of the observations for SDSS J1226$-$0006. Grism and filter: G300V$+$GG435.
SR collimator: 0.2\arcsec\ per pixel. Slitlets width: $1.0\arcsec$ (R = 400 at 5900 \AA).}
\label{refer_J1226}
\begin{flushleft}
\begin{tabular}{cccccc}
\hline 
\hline 
ID & Date & Seeing $[\arcsec]$ & Airmass & Weather \\
\hline 
1 & 16/05/2005 & 0.85 & 1.109 & Photometric\\
2 & 16/05/2005 & 0.84 & 1.099 & Photometric\\
3 & 16/05/2005 & 0.92 & 1.105 & Photometric\\
4 & 16/05/2005 & 1.02 & 1.125 & Photometric\\
5 & 16/05/2005 & 0.78 & 1.221 & Photometric\\
6 & 16/05/2005 & 0.89 & 1.299 & Photometric\\
7 & 16/05/2005 & 0.82 & 1.422 & Photometric\\
8 & 16/05/2005 & 0.88 & 1.576 & Photometric\\
\hline 
\end{tabular}
\end{flushleft}
\end{table}

\begin{table}[t!]
\caption[]{Journal of the observations for SDSS J1335$+$0118. Grism and filter: G300V$+$GG435.
HR collimator: 0.1\arcsec\ per pixel. Slitlets width: $1.0\arcsec$ (R = 210 at 5900 \AA).}
\label{refer_J1335}
\begin{flushleft}
\begin{tabular}{cccccc}
\hline 
\hline 
ID & Date & Seeing $[\arcsec]$ & Airmass & Weather \\
\hline 
1 & 03/02/2005 & 0.73 & 1.167 & Photometric\\
2 & 03/02/2005 & 0.71 & 1.133 & Photometric\\
3 & 03/03/2005 & 0.69 & 1.112 & Photometric\\
4 & 03/03/2005 & 0.77 & 1.123 & Photometric\\
5 & 03/03/2005 & 0.68 & 1.153 & Photometric\\
6 & 03/03/2005 & 0.62 & 1.198 & Photometric\\
\hline 
\end{tabular}
\end{flushleft}
\end{table}

\begin{table}[t!]
\caption[]{Journal of the observations for Q~1355$-$2257. Grism and filter: G300V$+$GG435.
HR collimator: 0.1\arcsec\ per pixel. Slitlets width: $1.0\arcsec$ (R = 210 at 5900 \AA).}
\label{refer_Q1355}
\begin{flushleft}
\begin{tabular}{cccccc}
\hline 
\hline 
ID & Date & Seeing $[\arcsec]$ & Airmass & Weather \\
\hline 
1 & 05/03/2005 & 0.68 & 1.016 & Photometric\\
2 & 05/03/2005 & 0.73 & 1.040 & Photometric\\
3 & 20/03/2005 & 0.63 & 1.038 & Photometric\\
4 & 20/03/2005 & 0.54 & 1.015 & Photometric\\
5 & 20/03/2005 & 0.57 & 1.105 & Photometric\\
6 & 20/03/2005 & 0.56 & 1.166 & Photometric\\
\hline 
\end{tabular}
\end{flushleft}
\end{table}

\begin{table}[t!]
\caption[]{Journal of observations for WFI J2033$-$4723. Grism and filter: G300V$+$GG435.
HR collimator: 0.1\arcsec\ per pixel. Slitlets width: $1.4\arcsec$ (R = 160 at 5900 \AA).}
\label{refer_WFI2033}
\begin{flushleft}
\begin{tabular}{cccccc}
\hline 
\hline 
ID & Date & Seeing $[\arcsec]$ & Airmass & Weather \\
\hline 
1 & 13/05/2005 & 0.50 & 1.256 & Light clouds\\
2 & 13/05/2005 & 0.58 & 1.198 & Light clouds\\
3 & 13/05/2005 & 0.60 & 1.148 & Light clouds\\
4 & 13/05/2005 & 0.48 & 1.117 & Light clouds\\
5 & 13/05/2005 & 0.53 & 1.095 & Light clouds\\
\hline 
\end{tabular}
\end{flushleft}
\end{table}

\section{VLT Spectroscopy}

\subsection{Observations}

We present observations for eight gravitationally  lensed quasars, 
in  order to  determine  the
redshift   of the  lensing   galaxy. The targets  are HE~0047$-$1756,
HE~0230$-$2130, HE~0435$-$1223,  SDSS~J1138$+$0314, SDSS~J1335$+$0118,
Q~1355$-$2257  (also known  as CTQ~327), 
SDSS~J1226$-$0006 and WFI~J2033$-$4723. 

Our spectroscopic observations are acquired with the FOcal Reducer and
low dispersion Spectrograph  (FORS1), mounted  on  the ESO Very  Large
Telescope.  Very importantly, all the observations  are carried out in
the MOS mode (Multi  Object Spectroscopy).   This strategy allows  the
simultaneous observation of the main target  and of several stars used
both as flux calibrators and as  reference PSF to spatially deconvolve
the data.

\begin{figure}[t!]
\begin{center}
\includegraphics[width=8cm]{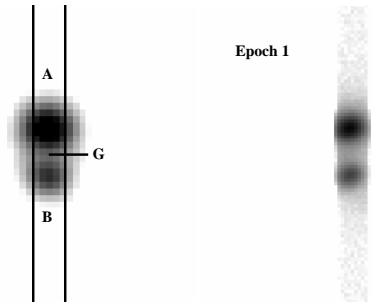}
\caption{HE~0047$-$1756.  Slit width: 1.0\arcsec. Mask PA : $10^{\circ}$}
\label{HE0047_slits}
\end{center}
\end{figure}

\begin{figure}[t!]
\begin{center}
\includegraphics[width=8cm]{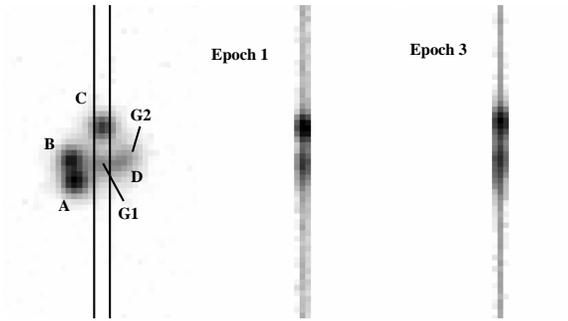}
\caption{HE~0230$-$2130.  Slit width: 0.5\arcsec. Mask PA : $-60^{\circ}$}
\label{HE0230_slits}
\end{center}
\end{figure}

\begin{figure}[t!]
\begin{center}
\includegraphics[width=8.cm]{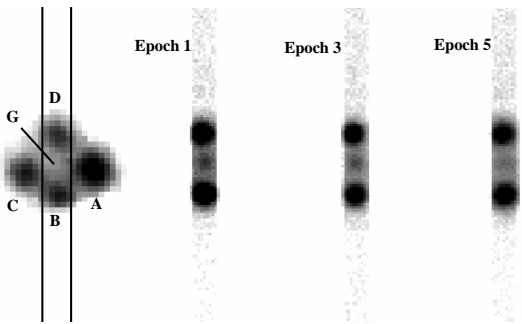}
\caption{HE~0435$-$1223. Slit width: 1.0\arcsec. Mask PA : $-164^{\circ}$}
\label{HE0435_slits}
\end{center}
\end{figure}

Most of our targets are observed with  the high-resolution collimator,
allowing us to observe  simultaneously eight objects over  a field of view of
$3.4\arcmin\times3.4\arcmin$   with a   pixel  scale  of $0.1\arcsec$.
However, because few suitable PSF stars are visible in the vicinity of
SDSS~J1226$-$0006  and HE~0230$-$2130,  the observations  for these two
objects  use the standard-resolution collimator, which  has a field of
view of $6.8\arcmin\times6.8\arcmin$ and a pixel size of 0.2\arcsec.

We use the GG435 order sorting filter in  combination with the G300V
grism for all objects, except HE~0230$-$2130 for which we use the G600R
grism. The G300V grism gives a useful wavelength range 
4450$~<\lambda<~$8650~\AA\ and a scale of $2.69$~\AA\     per 
pixel in the spectral direction. This setup has a spectral resolution
$R=\lambda/\Delta \lambda   \simeq  200$ at  the  central   wavelength
$\lambda=5900$~\AA\ for a $1.0\arcsec$ slit width in the case of the
high resolution collimator. The choice of  this
grism favors spectral coverage rather than spectral resolution as the
observations  are  aimed at  measuring   unknown lens  redshifts.  The
combination   of the  G600R grism  with   the GG435   filter used  for
HE~0230$-$2130 has a wavelength range of 
5250$~<\lambda<~$7450~\AA\  with a pixel scale of $1.08$~\AA\ in the spectral  direction.
This gives  a higher  spectral resolution of  $R \simeq  1200$  at the
central wavelength $\lambda=6270$~\AA\ for a slit width of $1.0\arcsec$
and with the standard resolution collimator.

\begin{figure}[t!]
\begin{center}
\includegraphics[width=8.cm]{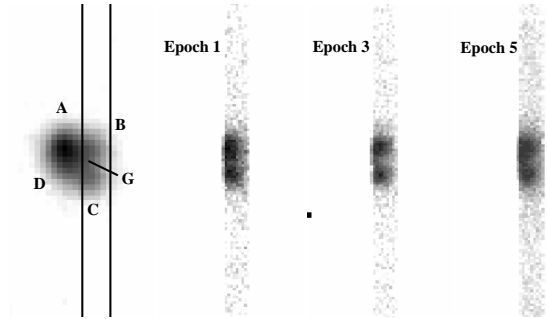}
\caption{SDSS~J1138$+$0314. Slit width: 1.0\arcsec. Mask PA : $-84^{\circ}$}
\label{J1138_slits}
\end{center}
\end{figure}

\begin{figure}[t!]
\begin{center}
\includegraphics[width=8.cm]{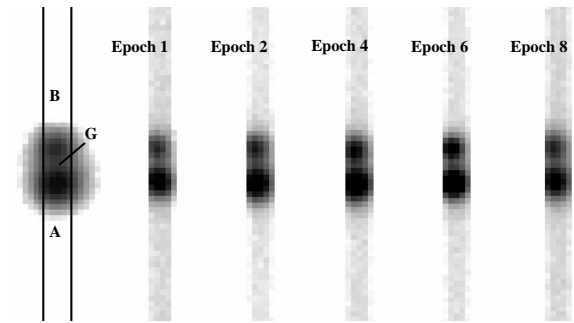}
\caption{SDSS~J1226$-$0006. Slit width: 1.0\arcsec. Mask PA : $-90^{\circ}$}
\label{J1226_slits}
\end{center}
\end{figure}

\begin{figure}[t!]
\begin{center}
\includegraphics[width=8.cm]{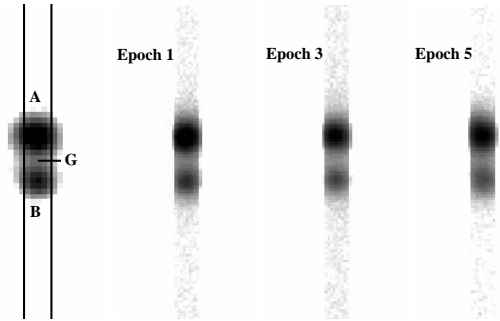}
\caption{SDSS~J1335$+$0118. Slit width: 1.0\arcsec. Mask PA : $43^{\circ}$}
\label{J1335_slits}
\end{center}
\end{figure}

\begin{figure}[t!]
\begin{center}
\includegraphics[width=8.cm]{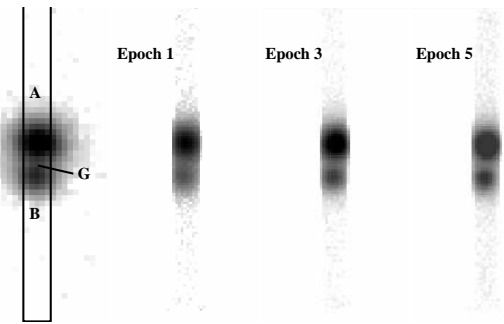}
\caption{Q~1355$-$2257. Slit width: 1.0\arcsec. Mask PA : $-78^{\circ}$}
\label{Q1355_slits}
\end{center}
\end{figure}

\begin{figure}[t!]
\begin{center}
\includegraphics[width=8.cm]{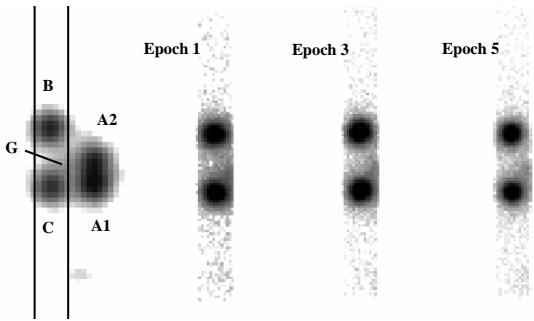}
\caption{WFI~J2033$-$4723. Slit width: 1.4\arcsec. Mask PA : $-82^{\circ}$}
\label{WFI2033_slits}
\end{center}
\end{figure}

We choose  slitlets of various widths,  depending on the brightness of
the  target and on the  configuration of the  quasar images.  The slit
width  is chosen so   that it  matches the  seeing  requested for  the
service-mode  observations     and  minimizes  lateral
contamination  by the quasar  images.  Our observing sequences consist
of a  short acquisition image, an ``image-through-slit'' check,  followed by
one or  two consecutive deep spectroscopic exposures.  All  individual exposures  for all
objects are 1400 s long.   The journals of  the observations are given
in  Tables~\ref{refer_HE0047} to \ref{refer_WFI2033}. The through-slit
images are displayed in Figs.~\ref{HE0047_slits} to \ref{refer_WFI2033},
where      the    epochs refer     to     the   exposure   numbers  in
Tables~\ref{refer_HE0047} to \ref{refer_WFI2033}.

For every  object we center  at least   three  slitlets on foreground
stars and one slitlet on the lensing  galaxy.  The mask is rotated to
a Position Angle that  avoids clipping of any  quasar image.  This  is
mandatory  to carry  out  spatial deconvolution of  the  spectra.  The
spectra of the  PSF stars are   also used to cross-calibrate the  flux
scale  as the  data  are  taken  at   different airmasses  and at
different dates.  Three spectrophotometric  standard stars are used to
carry  out the relative  flux  calibration, i.e.  EG~21, LTT~3218, and
LTT~6248.

In  addition  to our  own data,  we  retrieve VLT/FORS1  data from the
archive for HE~0230$-$2130.  These data were  acquired on 18 October
1999 for program 064.O-0259(A).   They used the
long  slit mode with the standard resolution collimator (0.2\arcsec\ per pixel),
the  G600R grism, and the order  sorting filter
GG435. The  useful  wavelength  range  is the  same  as  for the   MOS
observations, i.e., 5250$~<\lambda<~$7450~\AA\ with a pixel scale of 
$1.08$~\AA\ in the spectral direction.

\subsection{Reduction and Deconvolution}
 
We carry out the standard  bias subtraction and flat field  correction
of the spectra using IRAF\footnote{IRAF is distributed by the National
Optical Astronomy Observatories, which are operated by the Association
of Universities for  Research  in Astronomy, Inc.,  under  cooperative
agreement with  the National Science  Foundation.}.  The flat field is
created for each slitlet   from five dome  exposures using   cosmic ray
rejection.  It is  then normalized  by  averaging 45 lines along   the
spatial direction, rejecting the 20 highest  and 20 lowest pixels. The
result  is then block  replicated to  match the  physical size of  the
individual flat fields.

Wavelength calibration is obtained from numerous emission lines in the
spectrum of Helium-Argon lamps.  The  wavelength solution is fitted in
two dimensions to each  slitlet of the MOS  mask.  The fit uses a fifth
order Chebyshev  polynomial along the spectral  direction and a fourth
order Chebyshev   polynomial fit along  the  spatial  direction.  Each
spectrum is    interpolated  following this     fit  using a    cubic
interpolation.  This procedure  ensures that  the sky  lines  are well
aligned with the columns of the CCD after wavelength calibration.

The sky background is then removed by fitting and subtracting a second
order  Chebyshev polynomial in the spatial  direction to  the areas of
the spectrum that are not illuminated by the object.

Finally,  we  remove the cosmic rays   as follows. First  we shift the
spectra in order to  align them spatially (this   shift is only a  few
tenths of a  pixel). Second,  we create a  combined  spectrum for each
object from  all exposures, removing the  lower and higher pixels,
after applying   appropriate   flux scaling.  The  combined   spectrum
obtained in that way is cosmic ray cleaned and  is used as a reference
template to clean  the individual  spectra.   For  each MOS  mask  the
object  slitlet and the PSF slitlets  are  reduced exactly in the same
way.

A flux cross-calibration of  the spectra taken at different  airmasses
or dates is needed before combining  them into one final spectrum. This
is done efficiently using the spectra of the  PSF stars as references,
as described in Eigenbrod  et al.~(\cite{J0924}). The reference  stars
are assumed to  be non-variable  and  a ratio spectrum  is created for
each star, i.e. we divide the spectrum of the star by the spectrum of
the same star in the other exposures. This is done for at least two stars in
each mask   and we  check    that the response  curves   derived using
different  stars are compatible.   The dispersion between the response
curves obtained in that way is about 2\%. A mean correction curve is
then computed and applied to each two-dimensional spectrum. The 
individual spectra for each object are finally combined.
 
The  archive data  of  HE~0230$-$2130 consist of  one  single long slit
spectrum.   The bias subtraction, flatfielding, wavelength calibration
and background removal are done in exactly the same way as for the MOS
spectra.  The cosmic ray removal is done  using the IRAF packages for
single-image data.  The flux calibration  is done using three standard
stars:  G93$-$48, LTT~7987, LTT~9239, taken  on the same  night as the
science frame.

\begin{table}[t!]
\caption[]{Redshift values determined for the lensing galaxies in the eight
gravitational lenses. Only a tentative redshift is given for Q~1355$-$2257. 
See Section~\ref{he0230} for more details about
the second lens G2 in HE~0230$-$2130.}
\label{redshift}
\begin{flushleft}
\begin{tabular}{lc}
\hline 
\hline 
Object                  & z$_{\rm lens}$   \\
\hline 
HE~0047$-$1756          & $0.407 \pm 0.001$\\
HE~0230$-$2130, G1	& $0.523 \pm 0.001$\\
HE~0230$-$2130, G2	& $0.526 \pm 0.002$\\
HE~0435$-$1223    	& $0.454 \pm 0.001$\\
SDSS~J1138$+$0314       & $0.445 \pm 0.001$\\
SDSS~J1226$-$0006	& $0.517 \pm 0.001$\\
SDSS~J1335$+$0118       & $0.440 \pm 0.001$\\
Q~1355$-$2257   	& $0.701 (?)$\\
WFI~J2033$-$4723        & $0.661 \pm 0.001$\\  
\hline  	     
\end{tabular}	     
\end{flushleft}
\end{table}

\subsection{Deconvolution and extraction of the MOS spectra}

Even though the seeing values are good for most spectra, the lensing galaxy is
often close enough to the brighter quasar images to be affected by
significant  contamination from the wings of the PSF. For this reason,
the  spectral  version of the MCS deconvolution  algorithm (Magain et al. 
\cite{magain98},  Courbin et al.  \cite{courbin}) is  used in order to
separate  the spectrum of the  lensing galaxy from  the spectra of the
quasar  images.  The  MCS algorithm    uses  the spatial   information
contained  in the spectrum of a  reference PSF, which is obtained from
the slitlets positioned on  the isolated stars.  The final normalized
PSF   is a combination of at  least three different PSF spectra.  The
deconvolved  spectra are not only  sharpened in the spatial direction,
but also decomposed   into a ``point-source  channel'' containing  the
spectra of the  quasar  images, and a ``extended  channel'' containing
the spectra of everything  in the image  which is not a point-source, 
i.e. in this case the spectrum of the lensing galaxy.

\begin{figure}[t!]
\begin{center}
\includegraphics[width=8.7cm]{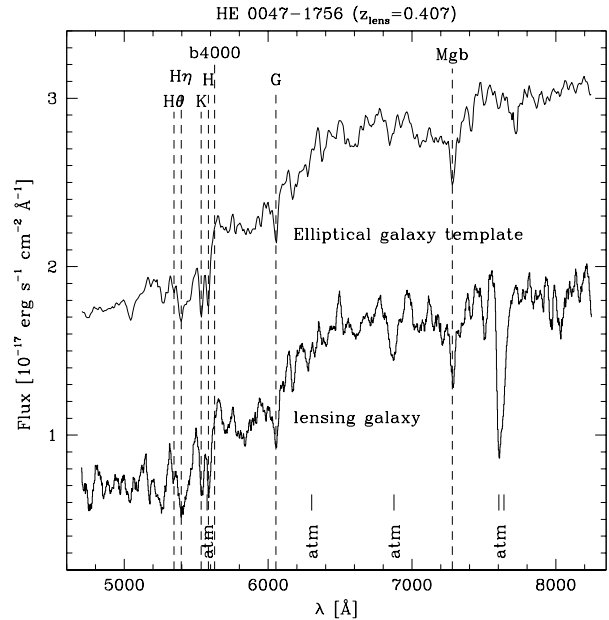}
\caption{Spectrum of the lens in HE~0047$-$1756. The
total integration time is 2800 s. The template spectrum of a redshifted 
elliptical galaxy  is shown for comparison (Kinney et al. \cite{kinney}).}
\label{HE0047_lens}
\end{center}
\end{figure}

\begin{figure}[t!]
\begin{center}
\includegraphics[width=8.7cm]{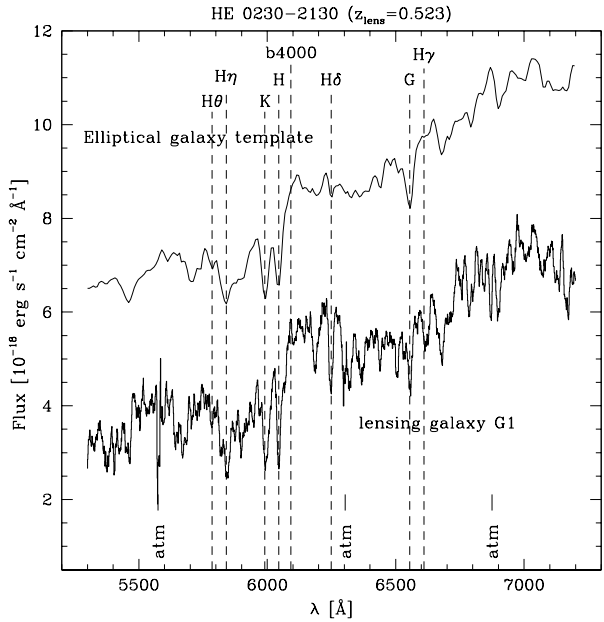}
\caption{Spectrum of the lensing galaxy G1 in HE~0230$-$2130, as obtained by
  combining the data  for the three epochs, i.e.  a  total integration time of
  4200~s.  A scaled version of the spectrum of quasar image C is subtracted
  to the spectrum of the lensing galaxy, in order to remove lateral contamination 
  by the quasar images A and B (see text).}
\label{HE0230_lens}
\end{center}
\end{figure}

\begin{figure}[t!]
\begin{center}
\includegraphics[width=8.7cm]{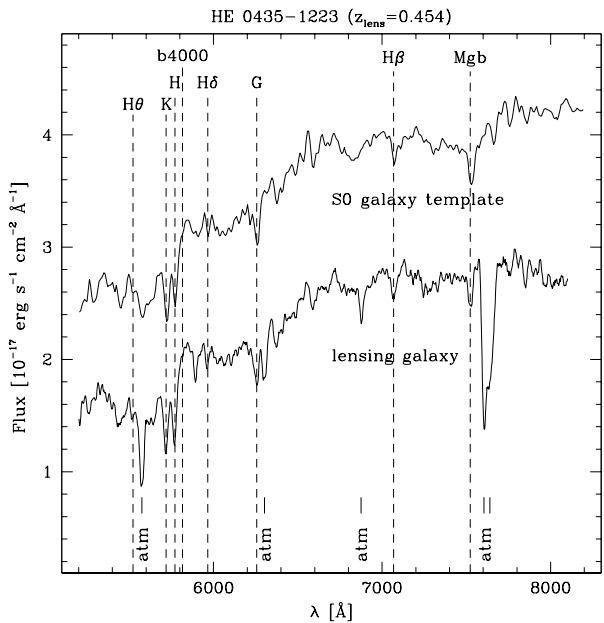}
\caption{Spectrum of the lens in HE~0435$-$1223. The
total integration time is 8400 s. The template spectrum of a redshifted S0 galaxy  is
shown (Kinney et al.  \cite{kinney}).}
\label{HE0435_lens}
\end{center}
\end{figure}

\begin{figure}[t!]
\begin{center}
\includegraphics[width=8.7cm]{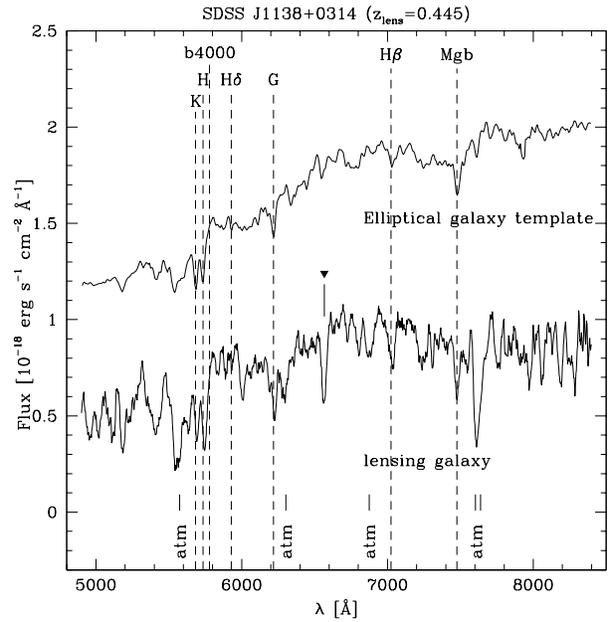}
\caption{Spectrum of the lens in SDSS~J1138$+$0314. The
total integration time is 7000 s. The absorption feature marked by 
the black triangle is probably residual light of the quasar images.
The C~III] emission of the quasar falls exactly at this wavelength.}
\label{SDSS1138_lens}
\end{center}
\end{figure}

\begin{figure}[t!]
\begin{center}
\includegraphics[width=8.7cm]{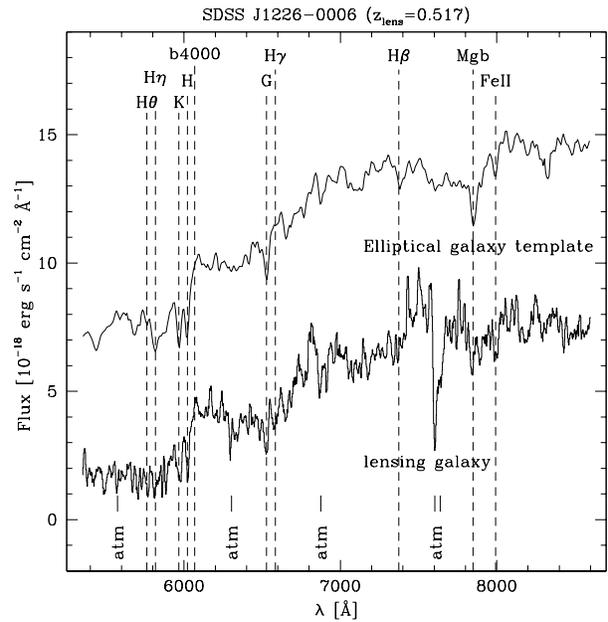}
\caption{Spectrum of the lens in SDSS~J1226$-$0006. The
total integration time is 11200 s.}
\label{SDSS1226_lens}
\end{center}
\end{figure}

\begin{figure}[t!]
\begin{center}
\includegraphics[width=8.7cm]{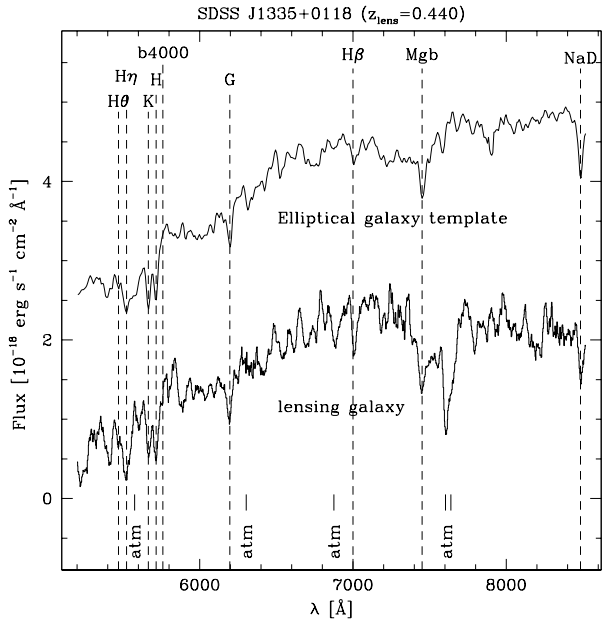}
\caption{Spectrum of the lens in SDSS J1335$+$0118. The 
total integration time is 8400 s.}
\label{J1335_lens}
\end{center}
\end{figure}

\begin{figure}[t!]
\begin{center}
\includegraphics[width=8.7cm]{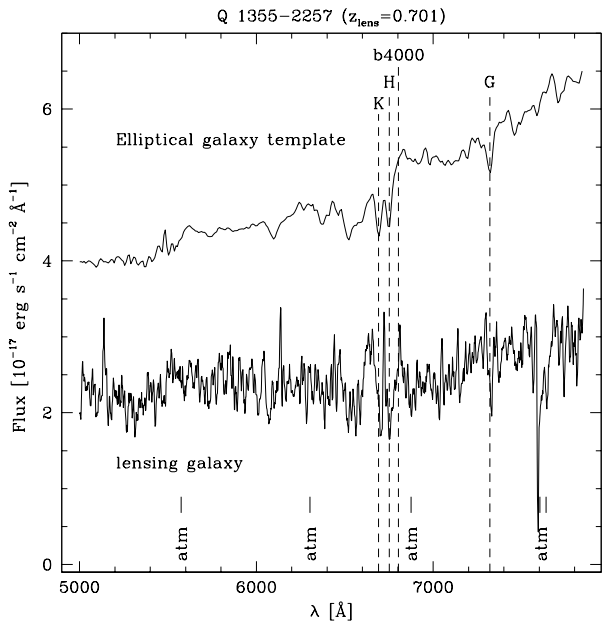}
\caption{Spectrum of the lens in Q~1355$-$2257. The 
total integration time is 8400 s. Given the low signal-to-noise of  
this spectrum we can not securely  determine the lens  redshift.}
\label{Q1355_lens}
\end{center}
\end{figure}

\begin{figure}[t!]
\begin{center}
\includegraphics[width=8.7cm]{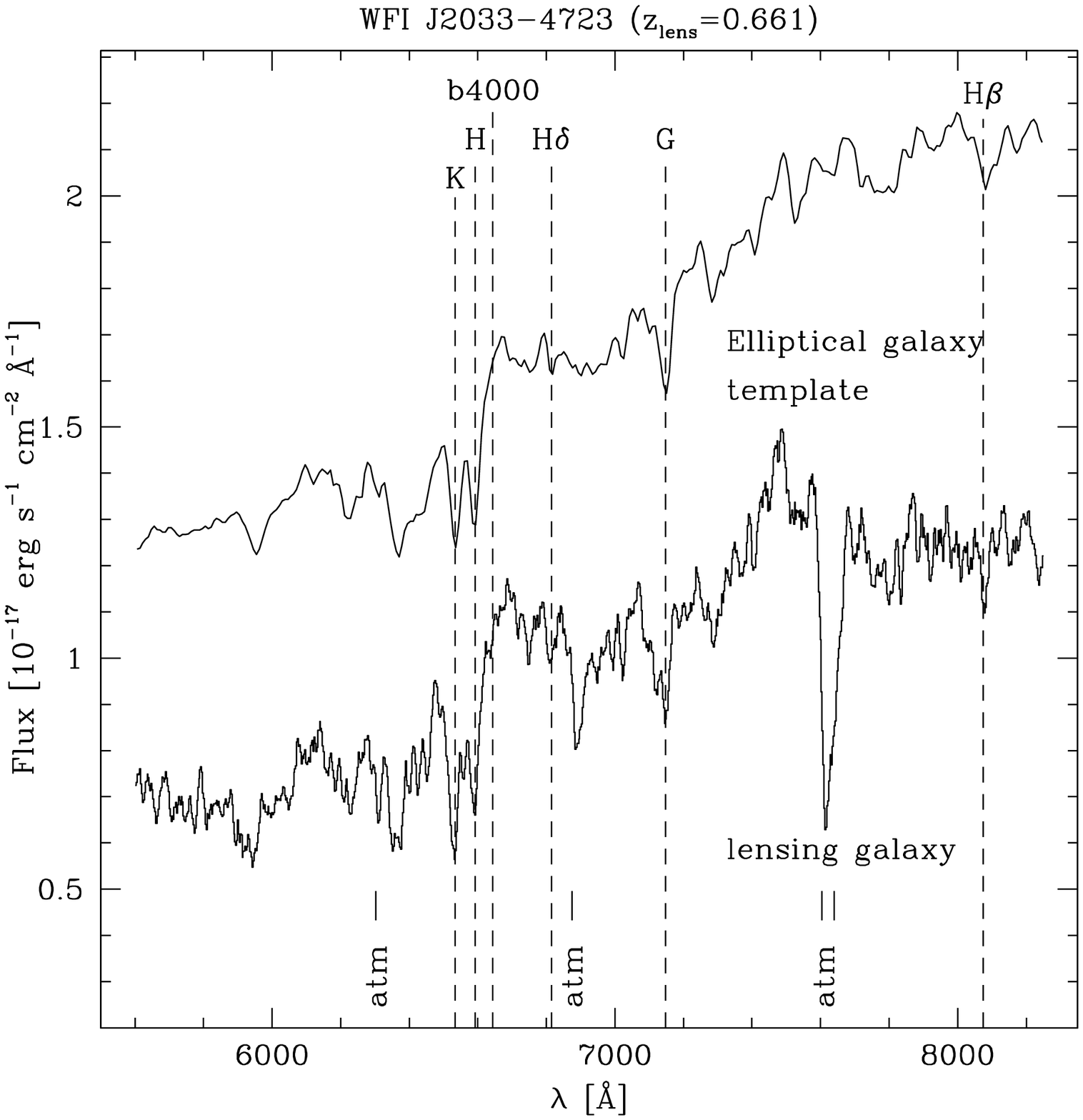}
\caption{Spectrum of the lens in  WFI~J2033$-$4723. The 
total integration time is 7000 s.}
\label{WFI2033_lens}
\end{center}
\end{figure}

The deconvolved   spectra of the lensing   galaxies are extracted and
smoothed    with   a   10~\AA\   box.     Fig.~\ref{HE0047_lens}  to
\ref{WFI2033_lens} display  the  final one-dimensional  spectra, where
the Ca~II H \& K absorption lines  are obvious, as well as the $4000$
\AA $\,$  Balmer break, and the G-band typical for  CH absorption. In some
cases, we  identify  a  few more features   that  are labeled  in  the
individual figures.   The identified lines   are used to  determine the
redshift of the lensing galaxies   given in Table~\ref{redshift}.   We
compute the 1-$\sigma$ error as the standard deviation between all the
measurements of  the individual lines.
The  absence of  emission  lines  in all spectra indicates that  the  
observed  lensing galaxies  are  gas-poor early-type galaxies. 

In most cases, no trace of the quasar  broad emission lines is seen in
the  spectrum of  the  lensing  galaxy,   indicative   of  an accurate
decomposition of the  data into the extended  (lens)  and point source
(quasar images) channels.  Only our VLT spectrum of the lensing galaxy
of HE~0230$-$2130  is  suffering from residuals  of  the  quasar broad
emission lines, probably due to  lateral contamination by images A and
B   of  the  quasar.  This   additional   source  of contamination  is
circumvented by subtracting a scaled version of the spectrum of quasar 
image C to the spectrum of the lensing  galaxy. The flux calibration of this
particular  lens might therefore be less  accurate  than for the other
objects.  Note, however, that  this   procedure  was only applied   to
HE~0230$-$2130.

\section{Notes on individual objects}

{\bf HE~0047$-$1756}:	
A doubly imaged    quasar  discovered  by   Wisotzki   et
al. (\cite{wiso04}). It has  a redshift of $z=1.67$ and a maximum
image separation of  1.44\arcsec. 
The redshift of the lensing
galaxy has  recently been measured by Ofek
et al.   (\cite{ofek}) at  \zl$=0.408$.  We confirm this result, 
with  \zl$=0.407 \pm  0.001$,  and  present  a  much  higher signal-to-noise
spectrum  in Fig.~\ref{HE0047_lens}. An  elliptical galaxy
template matches well the spectrum of the lens.

\begin{figure}[t!]
\begin{center}
\includegraphics[width=9.cm]{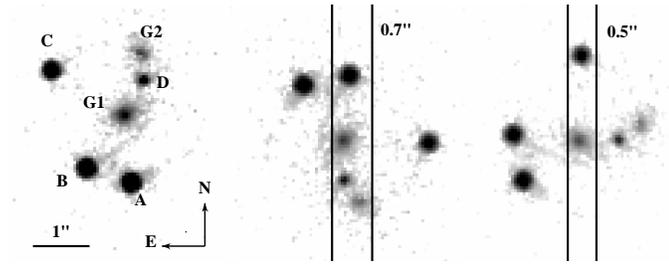}
\caption{\emph{Right:}  HST image of HE~0230$-$2130 taken with the 
WFPC2 instrument in the F814W filter. The pixel scale is $0.05 \arcsec$.
\emph{Middle:} same image but with a position angle of $-160^o$. We
show  in  overlay   a   $0.7\arcsec$ slit which    corresponds  to the
observational   setup used  to   take the  long-slit spectrum (program
064.O-0259(A) archive data). \emph{Left:}  for comparison, we show
the slit used for our MOS observations. The  position angle is $-60^o$
and the slit has a width of $0.5  \arcsec$. Negative angles are counted
clockwise from North.}
\label{HE0230_slits_HST}
\end{center}
\end{figure}

{\bf HE~0230$-$2130}:   This  quadruply  imaged  $z=2.162$ quasar  was
discovered by Wisotzki et al. (\cite{wiso99}).  It has a maximum image
separation of 2.15\arcsec\ and  two lensing galaxies. The main lensing
galaxy,  G1,  is located between  the  four quasar images.   A second,
fainter lens is located outside the area defined by the quasar images,
close to the faint quasar  image D (Fig.~\ref{HE0230_slits_HST}).  The
spectrum of G1 is shown in Fig.~\ref{HE0230_lens}.  Our redshift 
(\zl$=0.523 \pm 0.001$) is in very good agreement with the result 
\zl$=0.522$ of Ofek  
et al.~(\cite{ofek}). The spectrum matches well that of an early-type 
galaxy. The spectrum of the second lensing galaxy G2 is extracted from 
the archive long-slit spectrum presented in Section~\ref{he0230}. 

{\bf HE~0435$-$1223}:   
This quadruply   imaged    quasar,  discovered  by   Wisotzki   et
al. (\cite{wiso02}), has  a redshift of $z=1.689$ and has a maximum
image separation of  2.56\arcsec,  hence with little contamination  of
the lens spectrum  by the quasar  images.  The redshift of the lensing
galaxy has  already been measured  (Morgan et al.  \cite{morgan}, Ofek
et al.   \cite{ofek}) at  \zl$=0.455$.  We confirm this result, 
with  \zl$=0.454 \pm  0.001$,  and  present  a  much  higher signal-to-noise
spectrum  in Fig.~\ref{HE0435_lens}. A  S0 galaxy
template matches well the spectrum of the lens.

{\bf SDSS~J1138$+$0314}: This  quadruply imaged quasar was discovered 
in the course of the Sloan Digital Sky Survey (SDSS) by Burles et al. 
(\cite{burles}). This lensed quasar has a maximum image separation of 
$1.46 \arcsec$. We obtain high signal-to-noise spectra of the quasar
and determine its redshift $z=2.438$.
There is very little doubt that this system is lensed. The lensing 
galaxy is seen on archival HST/NICMOS  images. We measure
\zl$=0.445\pm0.001$ (Fig.~\ref{SDSS1138_lens}).

{\bf SDSS~J1226$-$0006}: A doubly imaged quasar at $z=1.120$ found in
the course of the Sloan Digital Sky Survey (SDSS) by Inada et al. 
(\cite{inada05}).  
This system  is doubtlessly lensed, as the lensing galaxy is   seen on
archival HST/ACS  images, between both quasar images, at only 
$0.4 \arcsec$ away from image A. 
We measure \zl$=0.516\pm0.001$ (Fig.~\ref{SDSS1226_lens}).
The spectrum of the lensing galaxy is well matched by the spectrum 
of an elliptical galaxy.

\begin{figure}[t!]
\begin{center}
\includegraphics[width=9.cm]{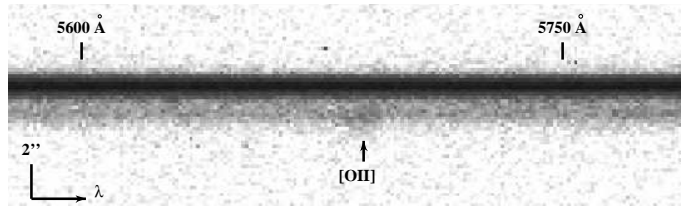}
\caption{Spectra of quasar image B and D of HE~0230$-$2130 (slit as in 
the middle  panel of Fig.~\ref{HE0230_slits_HST}.   A strong emission feature is seen
``below'' the  quasar  image   D. It   extends  well beyond  the  area
delimited by  the quasar images, and corresponds  to the [OII] emission
of galaxy G2.}
\label{HE0230_OII}
\end{center}
\end{figure}

{\bf SDSS~J1335$+$0118}:  A doubly  imaged  quasar with a 1.56\arcsec\
 separation, discovered by Oguri et al.  (\cite{oguri}). The quasar is
 at $z=1.57$ and the spectra of the quasar images show evidence of 
 strong absorption systems at lower redshifts. 
 Based  on the color of the  galaxy, Oguri et al.
 (\cite{oguri})  conclude that the lens galaxy  is  consistent with an
 early-type     galaxy  at      $z    \la    0.5$.    Our     spectrum
 (Fig.~\ref{J1335_lens})  is indeed that of  an early-type galaxy. Its
 redshift \zl$=0.440 \pm 0.001$ has not been measured before.

{\bf  Q~1355$-$2257 (CTQ~327)}:   This   doubly imaged  quasar   has a
redshift  of  $z=1.373$      and  was   discovered  by    Morgan    et
al.   (\cite{morgan03}).    The   quasar  images    are  separated  by
$1.23\arcsec$ and  the  redshift of  the   lensing galaxies has  been
estimated by Morgan et  al. (\cite{morgan03}) to  lie in  the redshift
range $0.4 <~$\zl$~< 0.6$. The extraction of the spectrum of the lensing
galaxy is particularly  difficult  because the  galaxy lies   at only
0.29\arcsec\ away from quasar image B and because  it is 7 magnitudes
fainter.  We show the deconvolved  spectrum of  the lensing galaxy  in
Fig.~\ref{Q1355_lens}.  Given the low signal-to-noise of  this spectrum we cannot
securely  determine the lens  redshift;  we can  only give a tentative
estimate of \zl$=0.701$.

{\bf  WFI~J2033$-$4723}:  Morgan et   al. (\cite{morgan04})
discovered this quadruply lensed quasar  with maximum image separation
of $2.53 \arcsec$ and  redshift $z=1.66$. Ofek  et al.  (\cite{ofek})
recently measured the  lens  redshift \zl$=0.658$. The  lensing galaxy
spectrum  shown   in  Fig.~\ref{WFI2033_lens}   is  matched   by  an
elliptical or S0   galaxy template with  a redshift  of \zl$=0.661 \pm
0.001$.  Our measured redshift  is compatible with but slighly higher
than the one reported by Ofek et al. (\cite{ofek}).

\begin{figure}[t!]
\begin{center}
\includegraphics[width=9.cm]{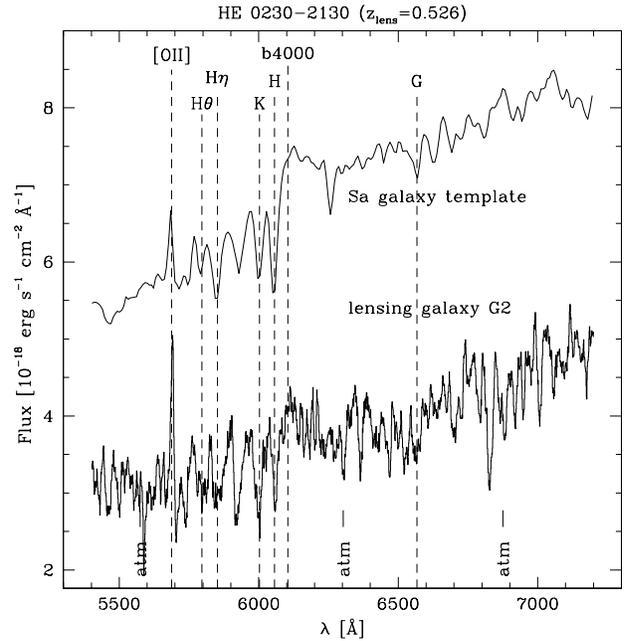}
\caption{Spectrum of the second lensing galaxy G2 in HE~0230$-$2130,
obtained from the FORS1 long-slit spectrum. The exposure time is 3000~s.
Note the prominent [OII]
emission lines, absent from the spectrum of G1 (Fig.~\ref{HE0230_lens}).}
\label{HE0230_lens_2}
\end{center}
\end{figure}

\section{The second lens in HE~0230$-$2130}
\label{he0230}

The long-slit archive  FORS1 spectrum of  HE~0230$-$2130 is taken with
the slit centered   on the quasar images   B  and D (middle  panel  in
Fig.~\ref{HE0230_slits_HST}).   The    two-dimensional  sky-subtracted
spectrum is   shown  in  Fig.~\ref{HE0230_OII}, where   a  hint  of an
emission line can already be seen at  the spatial location of lens G2
along the slit.	

As  no PSF star   is available the data  have  to be deconvolved in a
different way than the MOS   spectra.  We   proceed in  an   iterative
way. First, we cut the spectrum in two along the spectral direction and
determine a   first estimate of   the reference PSF spectrum  from the
brighter quasar image B directly.  During this first  step, the PSF is
constructed    from    the  upper     half    of   the    spectrum  in
Fig.~\ref{HE0230_OII}  containing the spectrum  of  image B.  We  then
deconvolve the original data, we take the extended-channel part of the
deconvolution, reconvolve it with  the approximate PSF and subtract it
from the data.  This provides us with a  spectrum of  the quasar image
that is  much less  affected  by  the lensing   galaxy.  
A second  PSF   is   constructed  from  these  new
lens-cleaned    spectra and a new    deconvolution  is carried out. We
perform this cycle  twice to  obtain  the final spectrum  presented in
Fig.~\ref{HE0230_lens_2}.  This spectrum has  a lower signal-to-noise than our MOS
spectra and less accurate flux calibration but it allows to measure well
the redshift  of   lens G2, which  displays   prominent [OII] emission,
contrary to  lens  G1. Fig.~\ref{HE0230_OII} shows   that the emission
comes from an object lying outside the region  delimited by the quasar
image. It therefore clearly corresponds to lens G2.

The redshift  of lens  G2  is \zl$=0.526\pm0.002$. It is  measured 
  using the [OII] emission, the Ca~II  H  \& K absorption
lines, the   G-band,  and  the hydrogen  H${\theta}$   and H${\eta}$
absorption lines.
The deconvolved spectrum  of G2 is shown in  Fig.~\ref{HE0230_lens_2}.
Although the flux calibration is not  optimal without a good knowledge
of the PSF,  the spectrum resembles that of  a Sa spiral galaxy.   

As can be seen in  the middle panel of Fig.~\ref{HE0230_slits_HST}, G2
is not well centered in the slit but lies at 0.4\arcsec\ away from the
slit center.  This small  {\it  spatial} misalignment of  the  object
within the slit mimics a  spectral shift of $\sim 2$  \AA\ to the red.
This translates into a redshift change of less than  $\Delta z = 0.0004$
at $6000$ \AA\ and  has  no effect on   the redshift determination  of
galaxy G2.  We   conclude  that the  observed difference  in  redshift
$\Delta z = 0.003$
between galaxy G1   and G2 is real.    It translates into  a  velocity
difference of  $\Delta v =  900 \pm 450 $~\kms\ typical for a galaxy 
group. HE~0230$-$2130 might therefore be lensed by a
group of physically related galaxies of which G1 and G2 are two of the
main members.

\section{Summary and Conclusions}

We present here the previously   unknown   redshifts of the lensing
galaxies in three  gravitationally lensed  quasars and confirm four  others,
already  presented in Ofek et al.~(\cite{ofek}).   We also measure the
redshift  of a  second lensing  galaxy   in HE~0230$-$2130 and  give a
tentative estimate of the lens redshift in Q~1355$-$2257.

The  MOS mode in which all  observations are taken and the subsequent
observation of  several PSF stars is crucial  to carry  out a reliable
decontamination   of the lens spectrum  by  those of the quasar images.
The  PSF stars are  also  used  to   carry out  a very  accurate  flux
calibration of the spectra.

Contrary to long-slit observations where   no PSF stars are  available
(Ofek et al.~\cite{ofek}), we  do not need  to iteratively remove  a 
scaled version of the quasar spectra from the data. Because microlensing
can produce significant differences between the  spectra of the quasar
images, such a procedure may   result in biased continuum slopes   and
wrong conclusions  about the presence of  dust in the lens.   For this
reason we fully  base our extraction on  a spatial decomposition using
independent PSF spectra.

We find that  all the lensing  galaxies  in our sample are  early-type
ellipticals or S0, except for the second lensing galaxy in
HE~0230$-$2130, which displays prominent [OII] emission. 
And we do  not  find any evidence  for significant
extinction by dust in their interstellar medium.

\begin{acknowledgements}
  The authors are very grateful  to the ESO  staff at Paranal for the
  particular care paid to the slit alignment necessary to perform the
  spectra deconvolutions. PM acknowledge financial support from PRODEX
  (Belgium). COSMOGRAIL is financially supported by the Swiss National
  Science Foundation (SNSF). The image shown in Fig.~\ref{HE0230_slits_HST}
  was obtained with the NASA/ESA
  Hubble Space Telescope (Program  \# 9744, PI: C.~S.  Kochanek) and
  extracted from the data  archives  at the Space Telescope  Science
  Institute,  which is operated   by the Association of Universities
  for	 Research    in Astronomy,    Inc.,    under  NASA  contract
  NAS~5-26555.
\end{acknowledgements}

\end{document}